\documentclass[aps,pre,twocolumn,superscriptaddress,showpacs]{revtex4-1}
\usepackage[dvips]{graphicx}
\usepackage{amssymb}
\usepackage{dcolumn}
\usepackage{bm}
\usepackage{amsmath}

\usepackage{color}

\begin{document}
 
\title{Reorientational solitons in nematic liquid crystals with modulated alignment}

\author{Filip A. Sala}

\affiliation{Warsaw University of Technology, Faculty of Physics, Warsaw 00-662, Poland}

\author{Noel F. Smyth}

\affiliation{School of Mathematics, University of Edinburgh, Edinburgh EH9 3FD, Scotland, U.K.}

\author{Urszula A. Laudyn}

\affiliation{Warsaw University of Technology, Faculty of Physics, Warsaw 00-662, Poland}

\author{\\Miros\l{}aw A. Karpierz}

\affiliation{Warsaw University of Technology, Faculty of Physics, Warsaw 00-662, Poland}

\author{Antonmaria A. Minzoni}

\affiliation{Universidad Nacional Aut\'onoma de M\'exico, Department of Mathematics and Mechanics, 
Instituto de Investigaci\'on en Matem\'aticas Aplic\'adas y Sistemas, M\'exico D.F. 01000, M\'exico.}

\author{Gaetano Assanto}

\affiliation{University of Rome ``Roma Tre'', NooEL--- Nonlinear Optics and OptoElectronics Lab, Rome 00146, Italy, and \\
Tampere University of Technology, Photonics Laboratory, Tampere FI-33101, Finland}


\begin{abstract}
In uniaxial soft matter with a reorientational nonlinearity, such as nematic liquid crystals, a  light 
beam in the extraordinary polarization walks off its wavevector  due to birefringence, 
while it undergoes self-focusing via an increase in refractive index and eventually forms a spatial soliton.  
Hereby the trajectory evolution of solitons in nematic liquid crystals--- nematicons--- 
in the presence of a linearly varying transverse orientation of the optic axis is analysed.  In this
study we use and compare two approaches: i) a slowly varying (adiabatic) approximation based on momentum conservation 
of the soliton in a Hamiltonian sense; ii) the Frank-Oseen elastic theory coupled with a fully vectorial and nonlinear 
beam propagation method.  The models provide comparable results in such a non-homogeneously oriented uniaxial medium and 
predict curved soliton paths with either monotonic or non-monotonic curvatures.  The 
minimal power needed to excite a solitary wave via reorientation remains essentially the same in both uniform and 
modulated cases. 
\end{abstract}

\pacs{42.65.Tg, 42.70.Df, 05.45.Yv}

\maketitle

This paper is dedicated to one of its coauthors, Professor Antonmaria (Tim) A. Minzoni, who prematurely passed away 
during its preparation. N.F.S. and G.A. remember Tim as a generous person of vast culture, a dear friend and an 
outstanding colleague.

\section{Introduction}

Nematic liquid crystals (NLCs) are anisotropic, typically uniaxial, soft matter with several peculiar properties.  
As suggested by the name, derived from the Greek, they consist of thread-like molecules and exhibit 
orientational but no spatial order \cite{khoo}.  The anisotropic molecules are in a fluid state, linked by elastic 
forces, and exhibit two refractive index eigenvalues, ordinary and extraordinary, for light polarized perpendicular 
or parallel to the optic axis, termed the molecular director and usually denoted by the unit vector 
$\vec{n}$.  The refractive index of extraordinary polarized light has a nonlinear optical dependence through the 
reorientational response: the electric field of the light beam induces dipoles in the 
NLC molecules, so that they tend to rotate towards the field vector to minimize the system energy until the elastic
response balances this electromechanical torque \cite{khoo}.   The resulting change in molecular orientation then
changes the extraordinary refractive index towards the largest eigenvalue, so that the beam undergoes self-focusing.  
When self-focusing compensates linear diffraction, a $(2+1)D$ solitary wave can form, often termed a nematicon 
\cite{PR,Wiley,karpierz6}.  Nematicons are non-diffracting solitary beams in nematic liquid crystals, confined by their 
own graded-index waveguides.  They have been extensively investigated over a number of years in many different scenarios, 
such as planar cells 
\cite{Peccianti:2000, Henninot:2002, Beeckman:2004, Assanto:2007, peccianti2008_mol_reorientation, Alberucci:2010_2, Alberucci_JOSAB, Izdebskaya:2010, Kwasny2012},
capillaries \cite{Warenghem_solitons1998, Warenghem:1998}, waveguides \cite{karpierz2002_physrevE,smyth2002} and bulk 
\cite{ISI:000393991400001}.  When the wavevector of the light beam and the molecular 
director are neither perpendicular nor parallel, the Poynting vector of the nematicon walks-off the wavevector at a 
finite angle owing to the tensorial nature of the dielectric susceptibility \cite{walk}.  Such an angular deviation of the 
energy flux depends on the refractive index eigenvalues, $n_{\parallel}$ and $n_{\perp}$ for electric fields parallel and 
perpendicular to the director, respectively, and the angle $\psi$ between the director and wavevector.  Nematicon walk-off 
can be exploited in optical devices, for instance, signal demultiplexers or routers 
\cite{Piccardi:2010, Piccardi:2010_2, Piccardi:2010_4, Piccardi:2010_5, Barboza:2011, Piccardi:2012_2, sala_jnopm2014}. 

In uniform NLCs, nematicons propagate along rectilinear trajectories along their Pointing vector. The corresponding 
graded index waveguides associated with these spatial solitons are therefore straight. Curved light induced waveguides 
have been investigated in NLCs by means of graded interfaces 
\cite{NPhys:2006,peccianti2008_mol_reorientation,Piccardi:2008,Barboza:2011,Piccardi:2012_2}, localized refractive 
index perturbations \cite{Pasquazi:2005,refract,Piccardi:2010_2}, interactions with boundaries 
\cite{Alberucci:2007_2,Alberucci_JOSAB,Izdebskaya:2010_2} as well as other nematicons 
\cite{Henninot:2002,Peccianti:2002_1,Fratalocchi:2007, Izdebskaya:2010,Jisha:2011}.
At variance with previous approaches, in this article we introduce and study curved reorientational spatial solitons 
as they propagate in nematic liquid crystals with a linearly varying orientation of the optic axis 
across the transverse coordinate in the principal plane (defined by director and wavevector). 
We consider nematicons excited in a planar cell of fixed (uniform) thickness, with upper and lower interfaces treated 
to ensure planar anchoring of the NLC molecules.  This geometry is radically different from those entailing spin-orbit 
interactions of light with matter \cite{Slussarenko:2016,Alberucci:2016,Alberucci:2017}, as the optic axis and the 
wavevector are not mutually orthogonal since the light beam is an extraordinary wave . 
As the molecular alignment varies across the sample, both the extraordinary refractive index and the birefringent 
walk-off vary as well.  These two variations determine the resulting trajectory of extraordinarily polarized
beams in the cell, including the path of self-confined nematicons.  
To investigate nematicon paths in a transversely modulated uniaxial we use two different approaches in the weakly nonlinear 
regime (i.e.\ power independent walk-off): (i) numerical solutions of the full governing Maxwell's equations employing a 
fully vectorial beam propagation method for the beam and the Frank-Oseen elastic theory for the NLC response \cite{PR}; 
(ii) an adiabatic (slowly varying) approximation to yield simplified forms of these equations, invoking momentum 
conservation \cite{PR,largeamp}.  The adiabatic approximation is based on the high nonlocality of the NLCs, which implies 
that the nonlinear response extends far beyond the transverse size of the optical wavepacket \cite{PR,conti2,JNOPM2016} 
and decouples the amplitude/width evolution of the beam from its trajectory \cite{ben,bennonlocal}.  In this study the 
background director angle is slowly varying, typically $0.002 \: \text{rad}$/$\mu \text{m}$ in a cell of width $200\:\mu \text{m}$, so 
that the nematicon trajectory can be determined by ``momentum conservation'', in the sense of invariances of the Lagrangian 
for the NLC equations.  The latter approach yields simple equations which have an exact solution and provides excellent 
agreement with the full numerical solutions, proving more than adequate to model beam evolution in non-uniform 
birefringent media.

\section{Geometry and governing equations}

\begin{figure}[h]
 \begin{center}
 \includegraphics[width=0.4\textwidth]{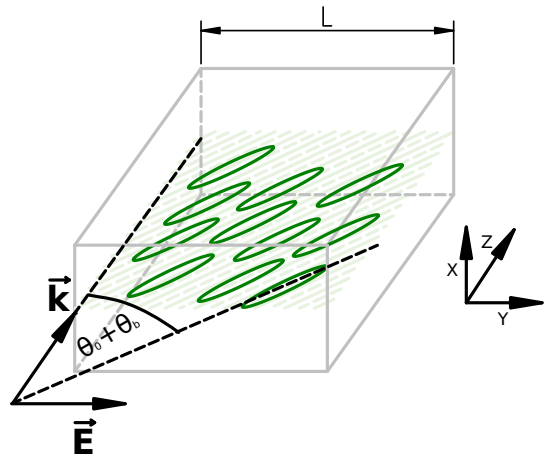}
 \end{center}
 \caption{(Color online) Sketch of the configuration and NLC alignment. The input Gaussian beam is linearly polarized 
 along $y$.}
 \label{fig:setup}
 \end{figure}

\begin{figure}[h]
 \begin{center}
 \includegraphics[width=0.5\textwidth]{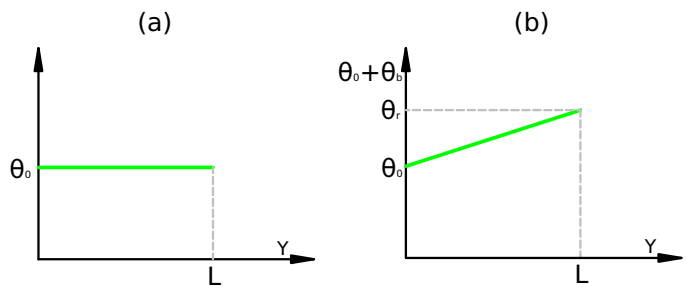}
 \end{center}
 \caption{(Color online) (a) Uniform and (b) linearly modulated anchoring conditions across $y$.} 
 \label{fig:anch_conditions}
 \end{figure}

We consider the propagation of a linearly polarized, coherent light beam in a cell filled with an undoped 
positive uniaxial nematic liquid crystal.  The extraordinary polarized beam is taken to initially propagate forward 
in the $z$ direction, with electric field $E$ oscillating in the $y$ transverse direction and $x$ 
completing the coordinate triad.  To eliminate the Fre\'edericksz threshold \cite{khoo} and maximize the nonlinear 
optical response \cite{Peccianti:2005}, the cell interfaces perpendicular to $x$ are rubbed so that the molecular 
director makes an angle $\theta_{0}$ with $z$ in the $(y,z)$ plane everywhere in the bulk owing to 
elastic interactions, as sketched in Fig.\ \ref{fig:setup}.  An additional $y$-dependent rotation $\theta_{b}(y)$ is 
given to the nematic director to modulate the uniaxial medium, as illustrated in Fig.\ \ref{fig:anch_conditions}(b).  
Due to the nonlinearity, the light beam can rotate the optic axis by an extra angle $\theta$, so that the director 
forms a total angle $\psi(y) = \theta_{0} + \theta_{b}(y) + \theta(y)$ to the $z$ axis in the $(y,z)$ plane 
\cite{Peccianti:2004,Peccianti:2005}.

\subsection{Beam propagation method and elastic theory}
\label{sec:elastic_theory}

One of the approaches used to study the nonlinear evolution of a light beam in nematic liquid crystals is the 
fully vectorial beam propagation method (FVBPM) \cite{ziogos2008} in conjunction with elastic theory based 
on the Frank-Oseen model for the NLC response \cite{khoo,frank,oseen}.  The FVBPM can be derived directly from 
Maxwell's equations \cite{sala2012_mclc, sala_mclc2012_Discrete_Diffraction}, 
considering harmonically oscillating electric and magnetic fields in an anisotropic dielectric
\begin{eqnarray}
\nonumber\frac{\partial H_z}{\partial y}-\frac{\partial H_y}{\partial z}&=&
i\omega\varepsilon_0\left(\varepsilon_{11}E_x+\varepsilon_{12}E_y+\varepsilon_{13}E_z\right), \\
\nonumber\frac{\partial H_x}{\partial z}-\frac{\partial H_z}{\partial x}&=&i\omega\varepsilon_0
\left(\varepsilon_{21}E_x+\varepsilon_{22}E_y+\varepsilon_{23}E_z\right), \\
\nonumber\frac{\partial H_y}{\partial x}-\frac{\partial H_x}{\partial y}&=&i\omega\varepsilon_0
\left(\varepsilon_{31}E_x+\varepsilon_{32}E_y+\varepsilon_{33}E_z\right), \\
\label{eq:fvbpm}
H_x&=&-\frac{1}{i\mu_0\omega}\left(\frac{\partial E_z}{\partial y}-\frac{\partial E_y}{\partial z}\right), \\
\nonumber H_y&=&-\frac{1}{i\mu_0\omega}\left(\frac{\partial E_x}{\partial z}-\frac{\partial E_z}{\partial x}\right), \\
\nonumber H_z&=&-\frac{1}{i\mu_0\omega}\left(\frac{\partial E_y}{\partial x}-\frac{\partial E_x}{\partial y}\right).
\end{eqnarray}
Here the complex amplitudes $\vec{E}$ and $\vec{H}$ are the electric and magnetic fields, respectively, $\varepsilon$ 
is the electric permittivity tensor, $\omega$ is the angular frequency and $\mu_0$ is the vacuum permeability.  These 
coupled partial differential equations can be solved numerically, as the $x$ and $y$ derivatives can be approximated
using standard central differences and the solution can be propagated forward along $z$ using a standard fourth-order 
Runge-Kutta scheme.
In this work the step size is chosen to be $dz=10\: \text{nm}$.  At the cell boundaries, reflective Dirichlet boundary 
conditions are imposed, so that ${\bf E} =0$ and ${\bf H} =0$ at the  NLC/glass interfaces.  The electric tensor in equations
(\ref{eq:fvbpm}) is
\begin{eqnarray}
\varepsilon=\left [
\begin{array}{ccc}
\varepsilon_{\perp} & 0 & 0\\
0 & \varepsilon_{\perp}+\Delta\varepsilon\sin^2\psi & \Delta\varepsilon \sin\psi\cos\psi\\
0 & \Delta \varepsilon \sin\psi \cos\psi  & \varepsilon_{\perp}+\Delta \varepsilon\cos^2\psi 
\end{array}
\right] ,
\end{eqnarray}
with $\Delta \varepsilon = n_{\parallel}^{2} - n_{\perp}^{2}$ the optical anisotropy.
These electromagnetic equations are coupled to the NLC response, given by the Frank-Oseen expression for 
the energy density in the non-chiral case \cite{khoo,frank,oseen}
\begin{eqnarray}
f& =& \frac{1}{2}K_{11}(\nabla \vec{n})^2+\frac{1}{2}K_{22}(\vec{n}\cdot (\nabla \times \vec{n}))^2 \nonumber \\
& & \mbox{} +\frac{1}{2}K_{33}(\vec{n}\times (\nabla \times \vec{n}))^2 
-\frac{1}{2}\varepsilon_0 \Delta \varepsilon (\vec{n}\cdot \vec{E})^2.
\label{eq:Frank_Oseen}
\end{eqnarray}
Here, $K_{11},K_{22},K_{33}$ are the Frank elastic constants for bend, twist and splay deformations of the molecular 
director $\vec{n}$, respectively \cite{PR}.  The equation for the NLC elastic response is obtained by taking variations 
of this free energy.  However, doing so results in a large system of equations \cite{sala_OptExpress2012}. 
To overcome this complexity, we note that in the examined configuration the molecular director and the electric field 
of the beam lie in the same (principal) plane $(y,z)$; hence, as nonlinear reorientation occurs in this same plane and  
the azimuthal components can be neglected, the director $\vec{n}$ can be expressed in polar coordinates 
$\vec{n}=[0,  \sin \psi,  \cos \psi]$.  Since the changes in molecular orientation along $z$  
are slow as compared with the wavelength of light, the derivatives with respect to $z$ can also be neglected.  In this 
approximation, the variations of the free energy (\ref{eq:Frank_Oseen}) yield the Euler-Lagrange equation
\begin{eqnarray}
\frac{\partial}{\partial x} \frac{\partial f}{\partial \psi_x}+\frac{\partial}{\partial y} 
\frac{\partial f}{\partial \psi_y}-\frac{\partial f}{\partial \psi}=0,
\label{eq:EulerLagrange}
\end{eqnarray}
leading to the director rotation in the form 
\begin{eqnarray}
& & K_{22}\frac{\partial^2\psi}{\partial x^2}+(K_{11}\cos^2\psi+K_{33}\sin^2\psi)\frac{\partial^2 \psi}{\partial y^2}
\nonumber \\
& & \mbox{} -\frac{1}{2}\sin 2\psi (K_{11}-K_{33})\left(\frac{\partial \psi}{\partial y} \right)^2 \nonumber \\
& & \mbox{} +\frac{\varepsilon_0 \Delta\varepsilon}{2}\left[2E_yE_z\cos 2\psi+\sin 2\psi (E_y^2-E_z^2)\right] =0.
\label{eq:case1}
\end{eqnarray}
Numerical solutions of this elliptic equation (\ref{eq:case1}) are found using successive over-relaxations (SOR) 
with relaxation parameter $\Omega=1.8$ \cite{SOR_book}.  When combined with the numerical solution of the electromagnetic 
model (\ref{eq:fvbpm}), solutions for beam propagation  in nematic liquid crystals with varying orientation can 
be obtained.  The director reorientation is recalculated after each $100$ nm of propagation;  after the first step in
$z$, the solution for $\psi$ is the initial guess for the SOR iterations, ensuring rapid convergence.
The accuracy of the method described above can be estimated from the ratio of total input and output powers, which 
should be unity because absorption is neglected and the boundary conditions are purely reflective.   Defining the relative 
error as $\eta = (P_{out}-P_{in})/ P_{in}$, we aim to achieve $|\eta| < 0.5\%$ for all the cases considered here.  
In this work, the typical cell dimensions (thickness $\times$ width $\times$ length) are 
$30\: \mu \text{m} \times 200\:\mu \text{m} \times 500\:\mu \text{m}$ and two simple anchoring conditions are analyzed, 
uniform and linearly varying, see Fig.\ \ref{fig:anch_conditions}.  For the sake of a realistic analysis, we choose the 
material parameters corresponding to the standard nematic liquid crystal 6CHBT, with Frank elastic constants 
$K_{11}=8.57\:\text{pN}$, $K_{22}=3.7\:\text{pN}$ and $K_{33}=9.51\:\text{pN}$ and indices $n_{\parallel} = 1.6335$ and 
$n_{\perp} = 1.4967$ at temperature T=$20^{o}C$ and wavelength $\lambda = 2\pi/k_{0} = 1.064\: \mu \text{m}$ 
\cite{walk,crc}. The input beam is Gaussian and $y$-polarized, with a full width half maximum 
$\text{FWHM}=7\:\mu \text{m}$ and power $1\:\text{mW}$. 

\subsection{Momentum conservation}
\label{s:momentum}

The full system (\ref{eq:fvbpm}) and (\ref{eq:case1}) governing the propagation of a light
beam in a non-uniform NLC cell is extensive and amenable to numerical solutions only.  However, these
equations can be simplified to yield a reduced system for which an adiabatic 
approximation applies based on the slow variation of the director orientation.  This adiabatic approximation
shows that the beam trajectory is determined by an overall ``momentum conservation'' (MC) equation.
This is not physical momentum, but momentum in the sense of the invariances of the Lagrangian in 
the reduced system.  Such reduction of the full system and the resulting momentum conservation
equation will now be derived.

The first approximation is that the imposed linear modulation $\theta_{b}$ in the director orientation is much smaller than 
the constant background $\theta_{0}$, $|\theta_{b}| \ll \theta_{0}$.  For the examples considered here,
typical values are $\theta_{0} = 45^{o}$ and maximum $|\theta_{b}|$ ranging from $5^{o}$ to $20^{o}$.
While the largest  $|\theta_{b}|$ is not strictly much smaller than $\theta_{0}$, nevertheless the asymptotic results
are found to be in good agreement with the numerical ones even at this upper limit.  As discussed in the previous 
section, we denote the additional nonlinear reorientation by $\theta$, 
so that the total pointwise orientation  is $\psi = \theta_{0} + \theta_{b} + \theta$.
In the paraxial, slowly varying envelope approximation, the equations (\ref{eq:fvbpm}) and (\ref{eq:case1}) governing 
the propagation of the light beam through the NLC can be reduced to \cite{conti2,PR,Wiley}
\begin{eqnarray}
& &  ik_{0}n_{e} \frac{\partial E_{y}}{\partial z} + 2ik_{0}n_{e} \Delta(\psi) \frac{\partial E_{y}}{\partial y}
+ \nabla^{2} E_{y} \nonumber \\
& & \mbox{} + k_{0}^{2} \left( n_{\perp}^{2} \cos^{2} \psi + n_{\parallel}^{2} \sin^{2} \psi \right.
 \nonumber \\
 & & \mbox{} \left. -  n_{\perp}^{2} \cos^{2} \theta_{0} - n_{\parallel}^{2} \sin^{2} \theta_{0} \right) E_{y} =
 0, \label{e:dimelect} \\
  & & K\nabla^{2} \psi + \frac{1}{4}\epsilon_{0} \Delta \epsilon |E_{y}|^{2} \sin 2\psi = 0.  
   \label{e:dimdirector}
\end{eqnarray}
As for the full equations of Section \ref{sec:elastic_theory}, $E_{y}$ is the complex valued envelope of 
the electric field of the beam, since in the paraxial approximation the components $E_{x}$ and $E_{z}$ are neglected.  
The Laplacian $\nabla^{2}$ is in the transverse $(x,y)$ plane.  In the single constant approximation,
the parameter $K$ is a scalar on the assumption that bend, splay and twist in the full director equation (\ref{eq:case1}) 
have comparable strengths.  The wavenumber $k_{0}$ of the input light beam is intended in vacuum and $n_{e}$ is the background 
extraordinary refractive index of the NLC \cite{PR,Wiley}
\begin{equation}
 n_{e}^{2}(\psi) = \frac{n_{\perp}^{2}n_{\parallel}^{2}}{n_{\parallel}^{2}\cos^{2}\psi
 + n_{\perp}^{2}\sin^{2} \psi},
 \label{e:n0}
\end{equation}
in the linear limit $\theta=0$.
The coefficient $\Delta$ is related to the birefringent walk-off angle $\delta$ of the extraordinary-wave beam, 
with $\tan \delta = \Delta$ in the $(y,z)$ plane, and is given by
\begin{equation}
 \Delta (\psi) = \frac{\Delta \epsilon \sin 2\psi}{\Delta \epsilon
 + 2n_{\perp}^{2} + \Delta \epsilon \cos 2\psi} .
 \label{e:delta}
\end{equation}
Throughout this work, despite the nonlinear dependence of $\Delta$ on the beam power through the reorientation 
$\theta$ \cite{Peccianti:2005_2,Piccardi:2010,Piccardi:2010_4}, we assume $\Delta=\Delta(\theta_0 + \theta_b)$ in 
the low power limit.  In the single elastic constant approximation, the director equations (\ref{eq:case1}) and 
(\ref{e:dimdirector}) differ by a factor of $1/2$ in the dipole term involving $\epsilon_{0} \Delta \epsilon$,
owing to definitions of the electric field based on either the maximum amplitude or the RMS (Root Mean Square) value.  In 
this context, this difference is equivalent to a rescaling of $K$, with the latter constant $K$ cancelling out in 
the adiabatic momentum conservation approximation.

The reduced  equations (\ref{e:dimelect}) and (\ref{e:dimdirector}) can be set in non-dimensional form 
via the variable and coordinate transformations
\begin{equation}
 x = WX, \quad y=WY, \quad z = BZ,\quad E_{y} = Au,
\label{e:scaledim}
\end{equation}
where
\begin{eqnarray}
& &  W = \frac{\lambda}{\pi \sqrt{\Delta \epsilon \sin2\theta_{0}}}, \quad
B = \frac{2n_{e}\lambda}{\pi \Delta \epsilon \sin 2\theta_{0}}, \nonumber \\
& & A^{2} = \frac{2P_{0}}{\pi \Gamma W^{2}}, \quad \Gamma = \frac{1}{2}\epsilon_{0}cn_{e}
\label{e:scales}
\end{eqnarray}
for a Gaussian input beam power of $P_{0}$ and wavelength $\lambda$ \cite{waveguide}.
With these non-dimensional variables,  Eqs. (\ref{e:dimelect}) and (\ref{e:dimdirector})
become
\begin{eqnarray}
 i \frac{\partial u}{\partial Z} + i \gamma \Delta(\theta_{0}+ \theta_{b})
 \frac{\partial u}{\partial Y} + \frac{1}{2} \nabla^{2} u \nonumber \\ 
 \mbox{} + 2\left( \theta_{0} 
 + \theta_{b} + \theta  \right) u & = & 0, \label{e:elec} \\
\nu \nabla^{2} \theta & = & - 2|u|^{2} \label{e:dir} .
\end{eqnarray}
In deriving these equations we assumed that the NLC director rotation
from $\theta_{0}$ is small, i.e., $|\theta_{b}| \ll \theta_{0}$, as discussed above.
We further assumed that the nonlinear response is small, with $|\theta| \ll \theta_{0}$.
The trigonometric functions in the dimensional equations (\ref{e:dimelect}) and 
(\ref{e:dimdirector}) have been expanded in Taylor series \cite{largeamp}.  The scaled parameters in 
these non-dimensional equations are
\begin{equation}
 \gamma = \frac{2n_{e}}{\sqrt{\Delta \epsilon \sin 2\theta_{0}}} \quad \mbox{and} \quad
\nu = \frac{8K}{\epsilon_{0}\Delta \epsilon A^{2}W^{2}\sin 2\theta_{0}}.
\label{e:scalevar}
\end{equation}

The equations (\ref{e:elec}) and (\ref{e:dir}) have the Lagrangian
formulation
\begin{eqnarray}
 L & = & i\left( u^{*}u_{Z} - uu_{Z}^{*} \right) + i\gamma \Delta(\theta_{0}
 +\theta_{b}) \left( u^{*}u_{Y} - uu_{Y}^{*} \right) \nonumber \\
 & & \mbox{} - |\nabla u|^{2} + 4\left( \theta_{0} + \theta_{b} + \theta \right)
 |u|^{2} - \nu |\nabla \theta |^{2},
 \label{e:lag}
\end{eqnarray}
where the $^{*}$ superscript denotes the complex conjugate.  Equations (\ref{e:elec}) and (\ref{e:dir}) 
have no general exact solitary wave, or nematicon, solution; the only known exact solutions are for 
specific, related values of the parameters \cite{mike}.  For this reason, variational and conservation law 
methods have proved to be useful to study nematicon evolution \cite{mike,tim}, as they give solutions in good agreement
with numerical and experimental results \cite{waveguide,mike,tim,wenjun}.  In particular, they provide 
accurate results for the refraction of nematicons due to variations in the dielectric constant 
\cite{wenjun,refract,wenjunrefract,scatter,wenjunvortex}.  Conservation laws based on the Lagrangian (\ref{e:lag}) 
are used below to determine the nematicon trajectory in a cell with an imposed linear modulation of the orientation 
angle $\theta_{0} + \theta_{b}$. 


The easiest way to obtain the approximate momentum conservation equations for Eqs. (\ref{e:elec}) and (\ref{e:dir}) is 
from the Lagrangian (\ref{e:lag}) \cite{bennonlocal,ben}.  We assume the general functional forms
\begin{equation}
 u = a g(\rho) e^{i\sigma + iV(Y-\xi)} \quad
\mbox{and} \quad \theta = \alpha g^{2}(\mu), 
 \label{e:trial}
\end{equation}
where
\begin{equation}
 \rho = \frac{\sqrt{X^{2} + (Y-\xi)^{2}}}{w}, \quad
 \mu = \frac{\sqrt{X^{2} + (Y-\xi)^{2}}}{\beta},
 \label{e:varab}
\end{equation}
for the nematicon and the director responses, respectively \cite{bennonlocal,ben}.  The actual beam profile $g$ is 
not specified, as the trajectory is found to be independent of this functional form \cite{ben}.  In response to the 
change in the NLC refractive index, the extraordinary wave beam undergoes refraction, as well as amplitude and width 
oscillations.  If the length scale of the refractive index change is larger than the beam width, the beam refraction 
decouples from the amplitude/width oscillations \cite{refract,scatter,bennonlocal}.  Consistent with this decoupling, 
the electric field amplitude $a$ and the width $w$ of the beam, the amplitude $\alpha$ and width $\beta$ of the director 
response can be taken as constant if just the beam trajectory is required.  Only the beam center position $\xi$ 
and (transverse) ``velocity'' $V$ are then taken to depend on $Z$, as well as the phase $\sigma$.  This approximation 
is equivalent to momentum conservation for the Lagrangian (\ref{e:lag}) \cite{newell}.  

Substituting the profile forms (\ref{e:trial}) into the Lagrangian (\ref{e:lag}) and 
averaging by integrating in $X$ and $Y$ from $-\infty$ to $\infty$  \cite{whitham}
gives the averaged Lagrangian \cite{tim}
\begin{eqnarray}
 {\cal L}_{m} & = & -2S_{2}\left( \sigma' -V\xi' \right) a^{2}w^{2}
 - S_{22}a^{2} \nonumber \\
 & & \mbox{} - S_{2}\left( V^{2} + 2VF_{1} - 4F \right) a^{2}w^{2} 
 + \frac{2A^{2}B^{2}\alpha \beta^{2} a^{2}w^{2}}{A^{2}\beta^{2} + B^{2}w^{2}} \nonumber \\
 & & \mbox{} - 4\nu S_{42}\alpha^{2} - 2qS_{4} \alpha^{2}\beta^{2} ,
 \label{e:avlag}
\end{eqnarray}
where primes denote differentiation with respect to $Z$.
Here $F$ and $F_{1}$, which determine the beam trajectory, are expressed by
\begin{eqnarray}
 F(\xi) & = & \frac{\int_{-\infty}^{\infty} \int_{-\infty}^{\infty} 
 \left( \theta_{0} + \theta_{b} \right) g^{2} \: dXdY}
 {\int_{-\infty}^{\infty} \int_{-\infty}^{\infty} g^{2} \: dXdY},
 \label{e:F} \\
 F_{1}(\xi) & = & \frac{\int_{-\infty}^{\infty} \int_{-\infty}^{\infty} 
 \gamma \Delta \left( \theta_{0} + \theta_{b} \right) g^{2} \: dXdY}
 {\int_{-\infty}^{\infty} \int_{-\infty}^{\infty} g^{2} \: dXdY}.
 \label{e:F1} 
 \end{eqnarray}
The integrals $S_{2}$, $S_{4}$ and $S_{22}$ and $S_{42}$ appearing in this averaged Lagrangian are
\begin{eqnarray}
& & S_{2} = \int_{0}^{\infty} \zeta g^{2}(\zeta) \: d\zeta, \quad
S_{22} = \int_{0}^{\infty} \zeta g'^{2}(\zeta) \: d\zeta, \nonumber \\
& & \label{e:sints} \\
& & S_{4} = \int_{0}^{\infty} \zeta g^{4}(\zeta) \: d\zeta, \quad
S_{42} = \frac{1}{4} \int_{0}^{\infty} \zeta \left[ \frac{d}{d\zeta} g^{2}(\zeta) 
\right]^{2} \: d\zeta. \nonumber
\end{eqnarray}
Taking variations of this averaged Lagrangian with respect to $\xi$ and $V$
yields the modulation equations
\begin{eqnarray}
 \frac{dV}{dZ} & = & 2\frac{dF}{d\xi} - V \frac{dF_{1}}{d\xi}, \label{e:varxi} \\
 \frac{d\xi}{dZ} & = & V + F_{1}, \label{e:vvar} 
\end{eqnarray}
which determine the beam trajectory.  Eq.\ (\ref{e:varxi}) is the momentum equation.  

A simple reduction of the trajectory Eqs.\ (\ref{e:varxi}) and (\ref{e:vvar})
can be carried out when the beam width is much less than the length scale for 
the variation of the refractive index, that is the length scale
of the variation of $\theta_{b}$ \cite{ben}.  For the examples in this work, 
$\theta_{b}' \sim 0.002 \:\text{rad}/\mu \text{m}$.  Hence, a length scale for the variation of $\theta_{b}$
is $500\:\mu \text{m}$, while the typical beam width is $7\:\mu \text{m}$.  The linear variation of the angle
$\theta_{b}$ from the background angle $\theta_{0}$ starts at $\theta_{b}=0$ at $Y=0$.  Since the beam is launched 
at the mid-section of the cell $Y=L/2$, where the total angle in the absence
of light is $\theta_{0} + \theta_{b}(L/2) = \theta_{m}$, it is more accurate to
expand the walk-off $\Delta$ in a Taylor series about $\theta_{m}$ rather than $\theta_{0}$.
If we set $\tilde{\theta}_{b} = \theta_{b} - \theta_{b}(L/2)$, the integrals $F$ 
(\ref{e:F}) and $F_{1}$ (\ref{e:F1}) can be approximated by
\begin{eqnarray}
 F(\xi) & \sim & \theta_{0} + \theta_{b}(\xi), \nonumber \\
 \quad F_{1}(\xi) & \sim &  
 \gamma \Delta(\theta_{0} + \theta_{b}(\xi)) \nonumber \\
 & = & \gamma \Delta(\theta_{m}) + 
 \gamma \Delta'(\theta_{m})\tilde{\theta}_{b}(\xi) + \ldots 
 \label{e:fexpand}
\end{eqnarray}
We note that $F_{1}$ has been further approximated by expanding  $\Delta$ in 
a Taylor series about $\theta_{0}$ on taking $|\theta_{b}| \ll \theta_{0}$, discussed above.  
With this simplification, the trajectory equations (\ref{e:varxi}) and (\ref{e:vvar}) become
\begin{eqnarray}
 \frac{dV}{dZ} & = & \left( 2 - V \gamma \Delta'(\theta_{m}) \right) \theta_{b}'(\xi) ,
 \label{e:veqn} \\
 \frac{d\xi}{dZ} & = & V + \gamma \Delta(\theta_{m}) + \gamma \Delta'(\theta_{m})\tilde{\theta}_{b}(\xi) . 
 \label{e:xieqn}
\end{eqnarray}

 \begin{figure}
 \begin{center}
 \includegraphics[width=0.45\textwidth]{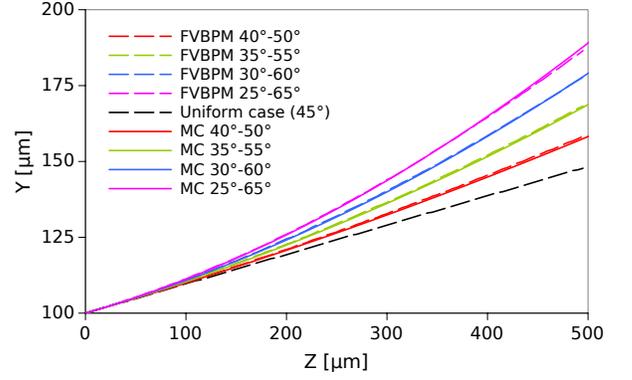}
 \end{center}
 \caption{(Color online) Numerical solutions of the NLC equations using FVBPM and elastic theory (dashed lines) 
 and  momentum conservation  (\ref{e:linsoln}) (solid lines), describing nematicon evolution in 
 a linearly modulated anchoring of NLCs in the cell .  Red line $\theta_{0}=40^{\circ}$ to $\theta_{r}=50^\circ$,  
 green line $\theta_{0}=35^{\circ}$ to $\theta_{r}=55^\circ$, blue line 
 $\theta_{0}=30^{\circ}$ to $\theta_{r}=60^\circ$ and pink line $\theta_{0}=25^{\circ}$ to $\theta_{r}=65^\circ$.} 
 \label{fig:comp}
 \end{figure}

The simplicity of the beam trajectory equations (\ref{e:veqn}) and (\ref{e:xieqn}) enables exact solutions
for simple angle modulations $\theta_{b}$.  The simplest is the linear case
\begin{equation}
 \theta_{b}(Y) = \frac{\theta_{r}}{L} Y,
 \label{e:thblinear}
\end{equation}
sketched in Fig.\ \ref{fig:anch_conditions}(b).
For this linear case, $\theta_{b}$ goes from $0$ at $Y=0$ to $\theta_{r}$ at $Y=L$.  
This variation of $\theta_{b}$ enables the momentum equations (\ref{e:veqn}) and (\ref{e:xieqn}) 
to be solved exactly and give the position of the beam center $\xi$ as
\begin{eqnarray}
 \xi & = & \left[ \xi_{0} + \frac{1+\gamma^{2}\Delta'(\theta_{m})\Delta(\theta_{m})}
{\gamma^{2}\Delta'^{2}(\theta_{m})\theta_{b}'} \right] e^{\gamma \Delta'(\theta_{m})\theta_{b}'Z} \nonumber \\
& & \mbox{} - \frac{2 + \gamma^{2}\Delta'(\theta_{m})\Delta(\theta_{m})}{\gamma^{2}\Delta'^{2}(\theta_{m})\theta_{b}'}
\nonumber \\
& & \mbox{} + \frac{1}{\gamma^{2} \Delta'^{2}(\theta_{m})\theta_{b}'} e^{-\gamma \Delta'(\theta_{m})\theta_{b}'Z} 
\label{e:linsoln}
\end{eqnarray}
as $\tilde{\theta}_{b}'=\theta'_{b}$ is a constant.  We assumed that the beam is launched at 
$\xi = \xi_{0}$ with $V=0$ at $Z=0$.  

Since $\theta_{b}$ is slowly varying, 
the trajectory solution given by Eq.\ (\ref{e:linsoln}) can be expanded in a Taylor series to yield
\begin{eqnarray}
 \xi & \sim & \left[ \xi_{0} + \gamma \Delta(\theta_{m})Z \right] \nonumber \\
& & \mbox{} + \left[ \xi_{0}\left( \gamma \Delta'(\theta_{m})\theta_{b}'Z
+ \frac{1}{2} \gamma^{2}\Delta'^{2}(\theta_{m})\theta_{b}'^{2}Z^{2}\right) + \right. \nonumber \\
& & \mbox{} \left. \left( 1 + \frac{1}{2}\gamma^{2} \Delta(\theta_{m})\Delta'(\theta_{m})\right)\theta_{b}' Z^{2} 
\right] + \ldots 
\label{e:smallthbp}
\end{eqnarray}
The first term in square brackets is the trajectory in a uniform NLC and the terms in the 
second set of square brackets are the correction due to a changing orientation.
For the examples hereby, $\theta_{b}' \sim 0.002 \:\text{rad}/\mu \text{m}$ and $\Delta ' \sim 0.05/\mu \text{m}$.  
So, to first order in small quantities
\begin{equation}
 \xi \sim \left[ \xi_{0} + \gamma \Delta(\theta_{m})Z \right] + \theta_{b}' Z^{2}
 \label{eq:trajectory_quadratic}
\end{equation}
as $\Delta'(\theta_{m})$ is small.  Hence, the trajectory is described by the term for a uniform medium and a 
quadratic correction; the walk-off change due to the varying background director orientation dominates the 
change in the nematicon trajectory.

To convert the non-dimensional solution (\ref{e:linsoln}) back to dimensional variables, the scalings
(\ref{e:scales}) are used.  In particular, for the $z$ scaling factor $B$, the angle for the extraordinary 
index (\ref{e:n0}) needs to be calculated.  The obvious choice is to use the uniform background angle $\theta_{0}$.  
However, while this leads to good agreement with the numerical solutions, near
exact agreement is obtained by using the total director angle $\theta_{0} + \theta_{b}$ in the 
absence of light.  The imposed component $\theta_{b}$ is not constant, but a slowly varying (linear) function of $Y$, 
as discussed above, so its local value can be used to transform
back to dimensional variables, consistent with a multiple scales analysis \cite{cole}.  This 
local variation in the scaling factor for $z$ gives a metric change 
in this coordinate, with a small, slowly varying alteration of the trajectory.  Nevertheless, the overall effect
of this small local change is significant over propagation distances of $500\:\mu \text{m}$ and larger.


\section{Results and discussion}
\label{s:results}


Figure \ref{fig:comp} shows a comparison of nematicon trajectories in the modulated NLC as given 
by the adiabatic momentum approximation (\ref{e:linsoln}) and by the FVBPM solution of the full 
system (\ref{eq:fvbpm}) and (\ref{eq:case1}).  The considered cell has 
a range of linear variations in the background director angle $\theta_{b}$ of the form (\ref{e:thblinear}).  Each
individual case, $\theta_{0} + \theta_{b}$, is indicated in the figure.  A Gaussian beam is launched at the center 
of the cell, with its trajectory becoming curved due to the non-uniform director alignment.  In a 
uniform medium the (straight) nematicon trajectory is determined solely by the walk-off, which leads to a rectilinear 
path in the $(y,z)$ plane.  For the modulated uniaxial medium, not only the walk-off changes due to the varying 
anchoring, but the phasefront of the wavepacket is also distorted as the dielectric properties are modified 
and the NLC behaves like a lens with an index distribution $n_e$ given by (\ref{e:n0}).  
Clearly, the momentum conservation approximation gives trajectories in close agreement with the numerical results.  
This validates the approximations made to arrive at the momentum conservation equations (\ref{e:veqn}) and 
(\ref{e:xieqn}), in particular the assumption that the beam trajectory is not influenced by its amplitude-width 
oscillations.  Furthermore, it shows how powerful such adiabatic approximations can be.  Nonetheless, the momentum 
result is a kinematic approximation and so does not give all the information for the evolving beam, whereas the 
full system (\ref{eq:fvbpm}) and (\ref{eq:case1}) can also provide the amplitude-width evolution.
A final point regarding Figure \ref{fig:comp} is that if the background 
angle for the extraordinary refractive index (\ref{e:n0}) in the $z$ scaling (\ref{e:scales}) was chosen
as $\theta_{0}$ rather than $\theta_{0} + \theta_{b}$, there would have been a noticeable difference
between the momentum conservation and numerical results.  The local variation of the propagation metric $z$
due to the modulated director angle in the absence of light, in fact, has a significant effect 
on beam propagation.

\begin{figure}
 \begin{center}
\includegraphics[width=0.48\textwidth]{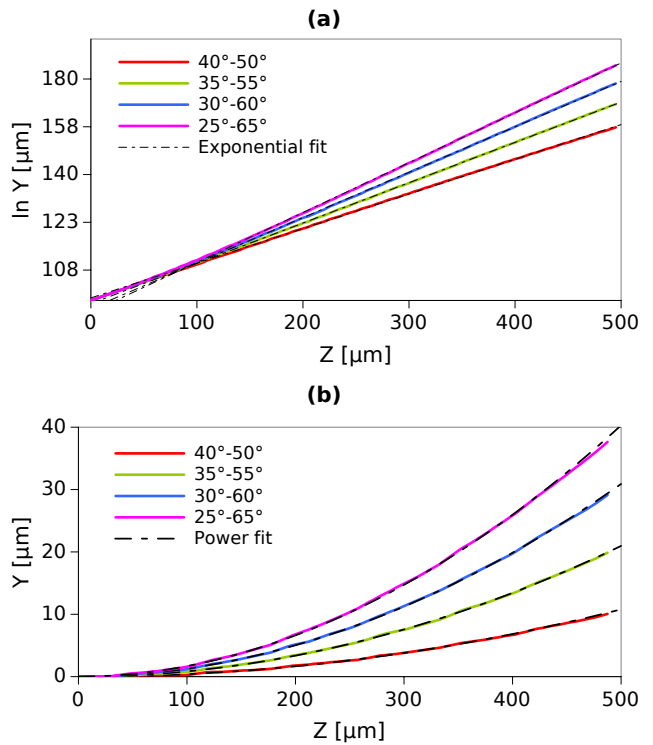}
 \end{center}
 \caption{FVBPM and elastic theory nematicon trajectory in a modulated cell: (a) logarithmic 
 scale with exponential fitting and (b) linear scale after subtracting the trajectory in a uniform 
 NLC ($\theta_{0}=45^\circ$, $\theta_{b}=0$). Red line $\theta_{0}=40^{\circ}$ to $\theta_{r}=50^\circ$,  
 green line $\theta_{0}=35^{\circ}$ to $\theta_{r}=55^\circ$, blue line $\theta_{0}=30^{\circ}$ to 
 $\theta_{r}=60^\circ$ and pink line $\theta_{0}=25^{\circ}$ to  $\theta_{r}=65^\circ$. Black dot dashed line: 
 (a) exponential fitting and (b) quadratic power fitting. Data fitted for $z>100\:\mu \text{m}$.}
 \label{fig:power_and_exp_fitting}
\end{figure} 

These results are further analyzed in Fig.\ \ref{fig:power_and_exp_fitting}(a).  The data is plotted
to a logarithmic scale with an exponential regression fitted through the numerical trajectories.  As $z$ increases 
the trajectories are well approximated by an exponential evolution, in agreement with the momentum conservation 
solution (\ref{e:linsoln}) as for large $z$ the decaying exponential is negligible and the growing exponential 
dominates.  Furthermore, when the rectilinear nematicon path in a uniform NLC is subtracted from the trajectory 
in the modulated case, the resulting beam position has a quadratic evolution in $z$, as shown in  
Fig.\ \ref{fig:power_and_exp_fitting} (b).  
These exponential and quadratic fittings of the trajectories are consistent with $\theta_{b}'$ and $\Delta'$
being small, as demonstrated by reducing the full trajectory (\ref{e:linsoln}) to the quadratic approximation
(\ref{eq:trajectory_quadratic}) via (\ref{e:smallthbp}).

\begin{figure}
 \begin{center}
 \includegraphics[width=0.48\textwidth]{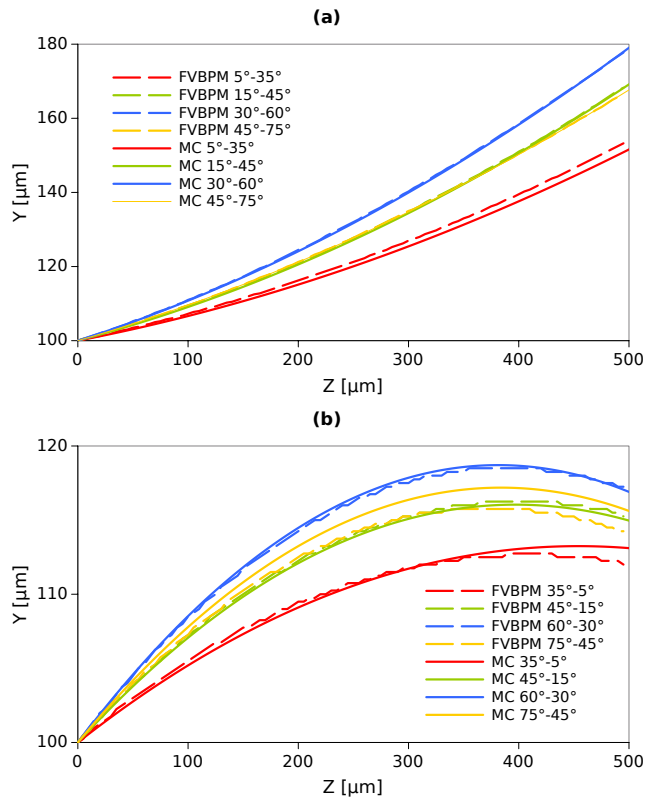}
 \end{center}
 \caption{Comparison of nematicon trajectories for (a) $\theta_r >\theta_0$ and (b) $\theta_r < \theta_0$.
 Numerical solutions of the nematic equations using FVBPM and elastic theory (dashed lines) and the momentum 
 conservation (\ref{e:linsoln}) (solid lines).  In all cases the rate of change is $30^\circ/200\:\mu \text{m}$.}
 \label{fig:comparison_slope30deg}
 \end{figure}

For a positive change of the anchoring conditions, i.e.\ $\theta_r >\theta_{0}$, walk-off and phase distortion both 
increase the beam deviation.  In the opposite case for which $\theta_r < \theta_{0}$ these two phenomena counteract.  
The influence of walk-off and phase change on the nematicon path was analyzed for the case of the director orientation 
changing by $30^\circ/200\:\mu \text{m}$, as shown in Fig.\ \ref{fig:comparison_slope30deg}.  When 
$\theta_r > \theta_{0}$ the beam bends strongly due to both the walk-off and phase distortions acting in the same 
direction, as illustrated in Fig.\ \ref{fig:comparison_slope30deg} (a).  The phase change is strongest at the launch position as the molecules 
are oriented at approximately $45^\circ$ there, so walk-off (given by (\ref{e:delta}) with $\psi = \theta_{0} + \theta_{b}$) 
is close to its maximum.  All the trajectories are monotonic and the beam transverse deviation increases with propagation 
distance.  As for the comparisons in Figure \ref{fig:comp}, the agreement between the momentum conservation and numerical
trajectories is near perfect, except for the lowest angle variation from $5^{o}$ to $35^{o}$, for which the 
agreement is still satisfactory.  In the latter case the initial director angle at the input is far from the walk-off 
maximum at $45^{o}$, so the trajectory bending is weak.  Small errors in the momentum approximation then become relevant.

In the opposite case $\theta_0 > \theta_r$ the walk-off and the phase change along the cell counteract, resulting 
in the solitary beam reversing its transverse velocity, as illustrated in the comparison of Fig.\ 
\ref{fig:comparison_slope30deg} (b).  The agreement between the momentum conservation and numerical trajectories
is nearly perfect, except for two noticeable cases.  The first is for the modulation from $35^{o}$ to $5^{o}$,  
opposite to what noted in the previous paragraph.  The reason for the disagreement is again the 
weak bending of the beam and the enhanced role of small errors in the momentum approximation.  The other case is the 
$75^{o}$ to $45^{o}$ modulation.  It can be seen from Fig. \ref{fig:comparison_slope30deg} (b) that 
as the range of $\theta_{b}$ varies the beam reaches a maximum deviation in $y$.  The $75^{o}$ to $45^{o}$ 
variation is just after this turning point.  As for the $35^{o}$ to $5^{o}$ case, small errors in the 
momentum conservation approximation can then result in large trajectory deviations, in particular errors
in the $\theta_{b}$ changes required for the maximum displacement in $y$.  

Finally, we note that comparable beam powers are needed to obtain nematicons in uniform and linearly modulated NLCs, 
as a $1\:\text{mW}$ input beam is sufficient to excite them in both cases, i.e.\ the rate of change in anchoring does not 
significantly modify the threshold power for reorientational solitons. 


\section{Conclusions}

We have studied the optical propagation of reorientational spatial solitons in nematic liquid crystals encompassing a 
transverse modulation of their optic axis (director) orientation.  Even in the simplest limit of a linear change 
in anchoring angle, as considered here, non-uniform walk-off and wavefront distortion determine a bending of the 
resulting nematicon trajectory, leading to curved paths and curved optical waveguides induced by light through reorientation. 
Based on comparisons with numerical solutions obtained by FVBPM and elastic theory for self-localized light beam propagation 
in non-uniform nematic liquid crystals, we found that ``momentum conservation'' is an excellent 
approximation for modelling soliton paths in highly nonlocal media.  It provides simple results for these
trajectories and a highly intuitive explanation for their evolution, at variance with the highly coupled form 
of the full governing equations.  While full numerical solutions can well 
describe nematicon evolution under generic conditions, the simplicity of the momentum 
conservation theory and its analytical solution speak in its favour for specific limits within the adiabatic 
category.  Due to the slow variation of the anchoring conditions, both models show that the nematicon trajectory 
can be described as propagation in a uniform medium with a quadratic correction.  Additionally, the power needed 
to excite reorientational solitons in either uniform or linearly non-uniform NLCs is comparable.  Further studies 
will investigate the role of longitudinal director modulations, as well as combinations of transverse and longitudinal 
changes, unveiling scenarios for the design of arbitrary nematicon paths and corresponding all-optical waveguides.


\section*{Acknowledgements}
F. A. Sala thanks the Faculty of Physics, Warsaw University of Technology, for a grant. U. A. L.  thanks the 
National Centre for Research and Development in Poland under the grant agreement LIDER/018/309/L-5/13/NCBR/2014. 
G. Assanto thanks the Academy of Finland for support through the Finland Distinguished Professor grant no.\ 282858.


\end{document}